\documentclass[twocolumn]{aastex631}
\usepackage{amsmath}
\usepackage{gensymb}

\shorttitle{Structure of the LMC Globular Cluster System from Proper Motions}
\shortauthors{Bennet et al.}
\graphicspath{{./}{figures/}}
\begin{document}

\title{Kinematic Structure of the Large Magellanic Cloud Globular Cluster System\\ from Gaia eDR3 and Hubble Space Telescope Proper Motions}

%\title{Template \aastex Article with Examples: 
%v6.3.1\footnote{Released on March, 1st, 2021}}

\correspondingauthor{Paul Bennet}
\email{pbennet@stsci.edu}

\author[0000-0001-8354-7279]{Paul Bennet}
\affiliation{Space Telescope Science Institute, 3700 San Martin Drive, Baltimore, MD 21218, USA}

\author[0000-0002-1212-2844]{Mayte Alfaro-Cuello}
\affiliation{Space Telescope Science Institute, 3700 San Martin Drive, Baltimore, MD 21218, USA}

\author[0000-0003-4922-5131]{Andr\'es del Pino}
\affiliation{Centro de Estudios de F\'isica del Cosmos de Arag\'on (CEFCA), Unidad Asociada al CSIC, Plaza San Juan 1, 44001, Teruel, Spain}
\affiliation{Space Telescope Science Institute, 3700 San Martin Drive, Baltimore, MD 21218, USA}

\author[0000-0002-1343-134X]{Laura L. Watkins}
\affiliation{AURA for the European Space Agency (ESA), ESA Office, Space Telescope Science Institute, 3700 San Martin Drive, Baltimore, MD 21218, USA}

\author[0000-0001-7827-7825]{Roeland P. van der Marel}
\affiliation{Space Telescope Science Institute, 3700 San Martin Drive, Baltimore, MD 21218, USA}
\affiliation{Center for Astrophysical Sciences, The William H. Miller III Department of Physics \& Astronomy, Johns Hopkins University, Baltimore, MD 21218, USA}

\author[0000-0001-8368-0221]{Sangmo Tony Sohn}
\affiliation{Space Telescope Science Institute, 3700 San Martin Drive, Baltimore, MD 21218, USA}

\begin{abstract}
We have determined bulk proper motions (PMs) for 31 LMC GCs from \textit{Gaia} eDR3 and \textit{Hubble Space Telescope} data using multiple independent analysis techniques. 
Combined with literature values for distances, line-of-sight velocities and existing bulk PMs, we extract full 6D phase-space information for 32 clusters, allowing us to examine the kinematics of the LMC GC system in detail. Except for two GCs (NGC~2159 and NGC~2210) for which high velocities suggest they are not long-term members of the LMC system, 
the data are consistent with a flattened configuration that rotates like the stellar disk. The one-dimensional velocity dispersions are of order 30 km~s$^{-1}$, similar to that of old stellar populations in the LMC disk. Similar to the case for Milky Way disk clusters, the velocity anisotropy is such that the dispersion is smallest in the azimuthal direction; however, alternative anisotropies cannot be ruled out due to distance uncertainties. The data are consistent with a single multi-dimensional Gaussian velocity distribution. Given the non-collisional nature of the LMC disk, this suggests that most, if not all, of the LMC GCs are formed by a single formation mechanism in the stellar disk, despite a significant spread in age and metallicity. Any accreted halo GC population is absent or far smaller in the LMC compared to the Milky Way. 
\end{abstract}


\section{Introduction} \label{sec:intro}

The Large Magellanic Cloud (LMC) is the most massive and closest known Milky Way (MW) satellite \citep[][]{McConnachie2012}. This galaxy has its own satellite - the Small Magellanic Cloud (SMC) - where signatures of interaction between them have been identified \citep{Belsa2012, Zivick2018}, as well as a stellar and HI bridge between them \citep{Hindman1963, Zivick2019}. This Magellanic system has also been associated with several ultra-faint dwarf galaxies (UFDs) that are believed to have been accreted by the MW alongside the LMC and SMC \citep[][]{Kallivayalil2018, Erkal2020, Patel2020}. 
Detailed studies of the LMC are important as they provide more information about a key component of the Local Group and the Magellanic system's potential impact on the MW.   
Therefore, the LMC offers an excellent case study for large dwarf galaxies and the wider Magellanic system provides one for how galaxy groups with a smaller central galaxy differ from those with masses comparable to the MW.

The globular clusters (GCs) of the LMC have been analyzed using line-of-sight (LoS) velocities for several decades. These studies generally found that the GCs are arranged in a disk-like structure with few, if any, showing halo-like kinematics \citep[for example][]{Schommer1992, Grocholski2006}. This is in contrast to the MW, which shows a combination of disk and halo GCs that differ in physical and kinematic properties \citep[see][for recent reviews]{Brodie2006, Vasiliev2021}. By studying GCs in detail around a host galaxy that has a significantly different GC population to the MW, we can begin to see how the mass and history of the host affects the distribution and properties of the GC system. This is particularly important as studies of extra-galactic GCs systems often include a color cut based on the MW's population \citep[][]{Peng2011, Beasley2016, Jones2021}, while in the LMC we find young GCs which are significantly bluer \citep[][]{Harris1996, Mackey2003}. This emphasizes the need to gain a deeper understanding about the LMC to inform future extra-galactic studies. 

In the \textit{Gaia} era, several studies of the proper motion (PM) of GCs have been published, using the second \textit{Gaia} Data Release (DR2), and third \textit{Gaia} early Data Release (eDR3) \citep[e.g.,][]{Helmi2018, Vasiliev2021}. 
However, these studies have focused on Galactic GCs. Examination of 15 LMC GCs using \textit{Gaia} DR2 data showed a possible dual population, with some clusters being limited to the disk and others more consistent with halo membership \citep[][]{Piatti2019}. This is controversial as it contradicts the findings of the LoS studies, which, as previously mentioned, show a single disk-like GC population. 
It would also produce a GC population in the LMC that is kinematically similar to the MW despite the significant differences in observational properties between the two galaxies' GCs, such as age and metallicity. 

In this work, we present a study using the proper motions from the \textit{Gaia} eDR3 \citep{Gaia_col_2020}, as well archival data from the Hubble Space Telescope (HST), of 42 GCs in the LMC, from this we find bulk PM results for  31 of these, over doubling the size of the previous sample. With existing literature LoS velocity and distances this allows us to construct full 6D information and analyze the cluster kinematics relative to the LMC. 

The structure of the paper is as follows: in \S \ref{sec:data} we discuss the data used, the sample of GCs examined and the methods used to determine cluster membership and compare the effectiveness of these methods. In \S \ref{sec:result} we discuss the results of these methods, the PM derived for each cluster, and the properties of each individual cluster and the population as a whole that can be derived from the PM measurements. Then in \S \ref{sec:discussion} we interpret these results and the implications and finally conclude in \S \ref{sec:conclusion}. 

%########################################################

\section{Data} \label{sec:data}

We begin with the LMC GC catalogues from \citet{Baumgardt2013} \& \citet{Piatti2018}. For all 42 identified GCs, we retrieve data within twice the reported cluster radii from the cluster centre from \textit{Gaia} eDR3 \citep{Gaia_col_2020}. 

Before beginning a detailed membership selection, we first make crude cuts to exclude obvious outliers and contaminants. We restrict the stellar search to magnitude G$_{mag}<19$, color $G_{BP}-G_{RP} > 0.6$ and errors in PM $\mu_{\alpha}$,$\mu_{\delta}$ $\leq0.5$~mas~yr$^{-1}$. This removes stars with magnitudes below the horizontal branch at the expected distance of the GCs and those with colors that are too blue to be likely cluster members, given the age and metallicity of the GCs. This is done to remove regions of the color-magnitude diagram (CMD) where we do not expect significant numbers of GC members, but that may contain LMC disk stars which could contaminate the sample.

To construct the 6D information (i.e. 3D position and 3D velocity information) for these clusters we also find the LoS velocities and distance estimates from the literature. We find LoS velocities for 37 of the 42 catalog clusters from a variety of sources; a full list of the LoS velocities along with references can be found in Table \ref{tab:prop}. 
From the full sample of 42 GCs in the initial catalogs we find distance estimates in the literature for 37 of these, which are also shown with references in Table \ref{tab:prop}. The uncertainties on the reported distances are in many cases large enough that the distances of most clusters are comparable. This presents a challenge in the interpretation of kinematics and is discussed further in \S \ref{subsec:spatial}.

\begin{deluxetable*}{c|cc|cc|cc}
\tablenum{1}
\tablecaption{Globular clusters properties\label{tab:prop}}
\tablewidth{0pt}
\tablehead{
\colhead{Cluster} & \colhead{RA} & \colhead{Dec} & \colhead{LoS Velocity} & \colhead{Source} & \colhead{Distance} & \colhead{Source} \\ 
\multicolumn1c{Name} & \multicolumn1c{(deg)} & \multicolumn1c{(deg)} & \multicolumn1c{(km~s$^{-1}$)} & \multicolumn1c{} & \multicolumn1c{(kpc)} & \multicolumn1c{} 
}
%\decimalcolnumbers
\startdata
NGC~1466 & 56.1371 & -71.6703 & 200.0$\pm$5.0 & \cite{Schommer1992} & 54.20$\pm$2.03 & \cite{Wagner-Kaiser2017} \\
NGC~1644 & 69.4125 & -66.1994 & 246.0$\pm$5.0 & \cite{Schommer1992} & 50.12$\pm$2.12 & \cite{Rosenfield2017} \\
NGC~1651 & 69.3796 & -70.5839 & 228.2$\pm$2.3 & \cite{Grocholski2006} & 48.08$\pm$0.67 & \cite{Goudfrooij2014} \\
NGC~1652 & 69.5917 & -68.6725 & 275.7$\pm$1.3 & \cite{Grocholski2006} & 49.66$\pm$1.16 & \cite{Milone2009} \\
NGC~1754 & 73.5753 & -70.4424 & 234.1$\pm$5.4 & \cite{Sharma2010} & 50.12$\pm$1.17 & \cite{Olsen1999} \\
NGC~1756 & 73.7083 & -69.2381 & - & - & - & - \\
NGC~1783 & 74.7833 & -65.9883 & 291.6$\pm$9.9 & \cite{Martocchia2021} & 49.20$\pm$0.91 & \cite{Goudfrooij2011} \\
NGC~1786 & 74.7811 & -67.7460 & 279.9$\pm$4.9 & \cite{Sharma2010} & 48.75$\pm$1.14 & \cite{Walker1988} \\
NGC~1805 & 75.5875 & -66.1117 & 390$\pm$10.0 & \cite{Fehrenbach1974} & 50.12$\pm$4.60 & \cite{Li2013} \\
NGC~1806 & 75.5458 & -67.9881 & 225.0$\pm$5.0 & \cite{Schommer1992} & 49.20$\pm$0.91 & \cite{Goudfrooij2011} \\
NGC~1818 & 76.0575 & -66.4339 & 311.1$\pm$3.9 & \cite{Marino2018} & 51.05$\pm$0.47 & \cite{Li2013} \\
NGC~1831 & 76.5725 & -64.9197 & 280.0$\pm$5.0 & \cite{Schommer1992} & 47.86$\pm$4.61 & \cite{Li2014} \\
NGC~1835 & 76.2884 & -69.4058 & 188.0$\pm$5.0 & \cite{Sharma2010} & 50.12$\pm$4.83 & \cite{Olsen1999} \\
NGC~1841 & 71.3496 & -83.9989 & 210.3$\pm$0.9 & \cite{Grocholski2006} & 52.00$\pm$1.21 & \cite{Wagner-Kaiser2017} \\
NGC~1866 & 78.4121 & -65.4644 & 298.4$\pm$0.4 & \cite{Mucciarelli2011} & 50.12$\pm$4.83 & \cite{Musella2016} \\
NGC~1868 & 78.6508 & -63.9539 & 283.0$\pm$3.0 & \cite{Olszewski1991} & 54.20$\pm$0.25 & \cite{Li2014} \\
NGC~1898 & 79.1891 & -69.6547 & 210.0$\pm$5.0 & \cite{Schommer1992} & 47.86$\pm$1.11 & \cite{Olsen1999} \\
NGC~1916 & 79.6578 & -69.4064 & 278.0$\pm$5.0 & \cite{Schommer1992} & 50.12$\pm$2.36 &  \cite{Olsen1999} \\
NGC~1928 & 80.2331 & -69.4764 & 249.6$\pm$12.8 & \cite{Piatti2018} & 50.35$\pm$1.41 & \cite{Fiorentino2011} \\
NGC~1939 & 80.3618 & -69.9497 & 258.8$\pm$7.4 & \cite{Piatti2018} & 49.66$\pm$3.80 & \cite{Mackey2004} \\
NGC~1978 & 82.1875 & -66.2361 & 292.0$\pm$2.5 & \cite{Hill2000} & 49.89$\pm$1.16 & \cite{Milone2009} \\
NGC~1987 & 81.8208 & -70.7356 & - & - & 47.42$\pm$4.83 & \cite{Goudfrooij2011} \\
NGC~2005 & 82.5354 & -69.7540 & 270.0$\pm$5.0 & \cite{Schommer1992} & 48.98$\pm$1.14 & \cite{Olsen1999} \\
NGC~2019 & 82.9853 & -70.1590 & 280.6$\pm$2.3 & \cite{Grocholski2006} & 50.12$\pm$1.17 & \cite{Olsen1999} \\
NGC~2031 & 83.4213 & -70.9869 & - & - & 48.31$\pm$4.83 & \cite{Testa2007} \\
NGC~2108 & 85.9833 & -69.1806 & - & - & 48.31$\pm$4.83 & \cite{Goudfrooij2011} \\
NGC~2156 & 89.4375 & -68.4606 & 285.0$\pm$17.0 & \cite{Ford1970} & - & - \\
NGC~2159 & 89.4875 & -68.6239 & 53.0$\pm$30.0 & \cite{Baird1986} & 52.48$\pm$2.47 & \cite{Elson1988} \\
NGC~2162 & 90.1267 & -63.7219 & 302.0$\pm$3.5 & \cite{Grocholski2006} & 52.48$\pm$0.24 & \cite{Pieres2016} \\
NGC~2173 & 89.4938 & -72.9778 & 237.4$\pm$0.7 & \cite{Grocholski2006} & 47.21$\pm$0.66 & \cite{Goudfrooij2014} \\
NGC~2190 & 90.2583 & -74.7258 & 260.0$\pm$3.0 & \cite{Olszewski1991} & 56.75$\pm$2.67 & - \\
NGC~2209 & 92.1450 & -73.8367 & 255.0$\pm$3.0 & \cite{Olszewski1991} & 48.53$\pm$2.05 & \cite{Kerber2007} \\
NGC~2210 & 92.8813 & -69.1214 & 343.0$\pm$5.0 & \cite{Schommer1992} & 50.58$\pm$0.94 & \cite{Wagner-Kaiser2017} \\
NGC~2231 & 95.1779 & -67.5194 & 277.6$\pm$1.4 & \cite{Grocholski2006} & 52.48$\pm$2.47 & \cite{Elson1988} \\
NGC~2249 & 96.4575 & -68.9203 & - & - & 45.08$\pm$1.26 & \cite{Kerber2007} \\
NGC~2257 & 97.5500 & -64.3267 & 301.6$\pm$0.8 & \cite{Grocholski2006} & 52.72$\pm$1.73 & \cite{Wagner-Kaiser2017} \\
Hodge~3 & 83.3350 & -68.1528 & 277.4$\pm$0.8 & \cite{Grocholski2006} & - & - \\
Hodge~4 & 83.1050 & -64.7364 & 310.8$\pm$1.9 & \cite{Grocholski2006} & 54.95$\pm$2.59 & \cite{Elson1988} \\
Hodge~7 & 87.5125 & -67.7181 & - & - & - & - \\
Hodge~11 & 93.5929 & -69.8472 & 245.1$\pm$1.0 & \cite{Grocholski2006} & 51.76$\pm$1.70 & \cite{Wagner-Kaiser2017} \\
Hodge~14 & 82.1637 & -73.6303 & 237.0$\pm$5.0 & \cite{Schommer1992} & - & - \\
Reticulum & 69.0475 & -58.8631 & 247.5$\pm$ 1.5 & \cite{Grocholski2006} & 51.76$\pm$1.45 & \cite{Wagner-Kaiser2017} \\
\enddata
%\tablecomments{}
\end{deluxetable*}

\subsection{Proper Motion Analysis} \label{subsec:select}

For consistency, we estimate the stellar membership selection and, therefore, the GC's bulk proper motions using two independent methods. These results were also then checked using the GaiaHub pipeline \citep{delPino2022}, which derives PMs from comparison of \textit{Gaia} and archival \textit{HST} positions, where applicable.

\subsubsection{Method 1} \label{subsec:M1}

In method 1, we retrieve the data and perform the membership selection of the GCs using \textsc{GetGaia}\footnote{https://github.com/AndresdPM/GetGaia} \citep[for more details see][]{delPino2021, Martinez-Garcia2021, GetGaia}.
We provide the names of the GCs to \textsc{GetGaia} so it retrieves the center of the GCs coordinates from Simbad for the search of the stars around a provided radius. We search for stars beyond several times the half-light radius of the GCs to have a significant non-members sample of stars for the stellar membership selection of the clusters. The number of non-member stars vary across the cluster sample; when the background is sparse, a larger radius is needed to ensure sufficient stars to characterise the background, but when the background is dense, a smaller radius suffices. We run \textsc{GetGaia} for all the GCs using a searching radius of 0.03 deg, and a second time, using a searching radius of 0.05 deg. We obtain consistent PMs for both runs and keep the results obtained with the searching radius that allows a better differentiation between member and non-member stars as observed in the CMD and vector point diagram.

After retrieving the data, this code applies a maximum likelihood approach to estimate the membership probability of each star using three parameters: the position of the star in the sky, the CMD, and PM-parallax space. We consider all the stars with membership probability $\geq50$\%. 
Afterwards, the code performs a refinement of the selection over the PM-parallax space using Gaussian Mixture Models. We use two Gaussian components for the PM and parallax clustering.
A detailed description of the membership selection from the background stars --stars in the LMC-- is presented by Watkins et al. (in prep) and in \citet{Martinez-Garcia2021}.

We estimate the bulk proper motions of the GCs using $3\sigma$ clipping for PM and parallax values. \textsc{GetGaia} provides the PM values with their corresponding errors, which are estimated in three ways: error-weighted mean, mean, and median absolute PM of the member stars. Amid some small differences, we find the three PM estimates to be consistent within $1\sigma$ for all the clusters in our sample. We decide to continue using the median, which gives more weight to the most numerous stars in the CMD and is more stable against outliers. The error provided by \textsc{GetGaia} for the median have to be interpreted as an upper limit, computed as $1.25*\sigma/\sqrt{N}$, with $N$ corresponding to the number of member stars. 

With this method, we obtain a reliable membership selection and bulk PMs for $22$ GCs from the $42$ in the catalogue, as listed in Table~\ref{tab:PM}. To illustrate the process, we show in the top row of Figure~\ref{fig:NGC1783} the CMD and vector point diagram for the proper motions for NGC~1783 obtained with this method, with the adopted cluster members. 

\begin{deluxetable*}{c|cc|c|cc|c|cc|c}
\tablenum{2}
\tablecaption{Globular cluster proper motions\label{tab:PM}}
\tablewidth{0pt}
\tablehead{
\colhead{Cluster} & \colhead{Method 1} & \colhead{Method 1} & \colhead{Method 1} & \colhead{Method 2} & \colhead{Method 2} & \colhead{Method 2} & \colhead{GaiaHub} & \colhead{GaiaHub} & \colhead{GaiaHub} \\
\multicolumn1c{Name} & \multicolumn1c{PM R.A.} & \multicolumn1c{PM Decl.} & \multicolumn1c{Star Count} & \multicolumn1c{PM R.A.} & \multicolumn1c{PM Decl.} & \multicolumn1c{Star Count} &  \multicolumn1c{PM R.A.} & \multicolumn1c{PM Decl.} & \multicolumn1c{Star Count}
}
%\decimalcolnumbers 
\startdata
NGC~1466 & 1.659$\pm$0.050 & -0.668$\pm$0.061 & 96 & \textbf{1.677$\pm$0.033} & \textbf{-0.679$\pm$0.024} & 122  & - & - & -  \\
NGC~1644 & \textbf{1.708$\pm$0.123} & \textbf{-0.304$\pm$0.131} & 38 & - & - & -  & 1.979$\pm$0.004 & -0.044$\pm$0.004 & 199  \\
NGC~1651 & \textbf{1.941$\pm$0.178} & \textbf{-0.325$\pm$0.158} & 51 & - & - & -  & - & - & -  \\
NGC~1652 & \textbf{1.783$\pm$0.325} & \textbf{-0.667$\pm$0.339} & 25 & - & - & -  &  1.846$\pm$0.003 & -0.377$\pm$0.003 & 380  \\
%NGC~1754 & - & - & - & - & - & - \\
NGC~1756 & - & - & - & - & - & -  & 1.859$\pm$0.010 & -0.049$\pm$0.010 & 1437  \\
NGC~1783 & \textbf{1.626$\pm$0.043} & \textbf{-0.019$\pm$0.050} & 426 & 1.589$\pm$0.150 & 0.033$\pm$0.126 & 42 &  1.645$\pm$0.010 & 0.024$\pm$0.010 & 1782 \\
NGC~1786 & \textbf{1.953$\pm$0.098} & \textbf{0.054$\pm$0.159} & 80 & - & - & - & - & - & -  \\
%NGC~1805 & - & - & - & - & - & - \\
NGC~1806 & \textbf{1.857$\pm$0.064} & \textbf{-0.043$\pm$0.067} & 207 & - & -  & - & 1.812$\pm$0.009 & -0.013$\pm$0.009 & 2282  \\
NGC~1818 & - & - & - & \textbf{1.526$\pm$0.032} & \textbf{0.126$\pm$0.048} & 56 & -  & - & -  \\
NGC~1831 & \textbf{1.681$\pm$0.045} & \textbf{-0.001$\pm$0.061} & 234 & 1.802$\pm$0.106 & 0.023$\pm$0.030 & 99 & - & - & - \\
%NGC~1835 & - & - & - & - & - & - \\
NGC~1841 & \textbf{1.966$\pm$0.018} & \textbf{-0.004$\pm$0.017} & 96 & 1.978$\pm$0.081 & 0.081$\pm$0.065 & 248 & - & - & - \\
NGC~1866 & - & - & - & \textbf{1.567$\pm$0.047} & \textbf{0.193$\pm$0.072} & 184 & - & - & -  \\
NGC~1868 & \textbf{1.619$\pm$0.167} & \textbf{0.202$\pm$0.257} & 52 & 1.627$\pm$0.262 & 0.078$\pm$0.190 & 65 & - & - & - \\
%NGC~1898 & - & - & - & - & - & - \\
%NGC~1916 & - & - & - & - & - & - \\
NGC~1928 & - & - & - & - & - & - & \textbf{1.870$\pm$0.01} & \textbf{0.305$\pm$0.01} & 3032  \\
NGC~1939 & - & - & - & - & - & - & \textbf{1.972$\pm$0.01} & \textbf{0.300$\pm$0.01} & 2828  \\
NGC~1978 & 1\textbf{.785$\pm$0.048} & \textbf{0.401$\pm$0.058} & 415 & - & - & -  & 1.741$\pm$0.01 & 0.454$\pm$0.01 & 1815  \\
NGC~1987 & - & - & - & - & - & - & 2.023$\pm$0.009 & 0.402$\pm$0.009 & 2189  \\
%NGC~2005 & - & - & - & - & - & - \\
%NGC~2019 & - & - & - & - & - & - \\
%NGC~2031 & - & - & - & - & - & - \\
NGC~2108 & - & - & - & - & - & - & 1.644$\pm$0.01 & 0.740$\pm$0.01 & 1529  \\
%NGC~2156 & - & - & - & - & - & - \\
NGC~2159 & - & - & - & \textbf{1.543$\pm$0.075} & \textbf{0.884$\pm$0.031} & 34 & - & - & - \\
NGC~2162 & \textbf{1.444$\pm$0.084} &\textbf{ 0.847$\pm$0.076} & 58 & 1.457$\pm$0.075 & 0.914$\pm$0.182 & 86 & - & - & - \\
NGC~2173 & 1.949$\pm$0.059 & 0.882$\pm$0.097 & 94 & \textbf{1.940$\pm$0.043} & \textbf{0.876$\pm$0.031} & 54 & - & - & - \\
NGC~2190 & 1.933$\pm$0.139 & 0.957$\pm$0.177 & 29 & \textbf{1.887$\pm$0.052} & \textbf{0.881$\pm$0.056} & 73 & - & - & - \\
NGC~2209 & 1.997$\pm$0.268 & 0.859$\pm$0.158 & 35 & \textbf{1.888$\pm$0.154} & \textbf{0.957$\pm$0.171} & 89 & - & - & - \\
NGC~2210 & 1.554$\pm$0.072 & 1.360$\pm$0.076 & 105 & \textbf{1.550$\pm$0.066} & \textbf{1.346$\pm$0.025} & 65 & - & - & - \\
NGC~2231 & 1.569$\pm$0.161 & 1.137$\pm$0.142 & 53 & \textbf{1.701$\pm$0.046} & \textbf{1.145$\pm$0.114} & 35 & - & - & - \\
NGC~2249 & 1.947$\pm$0.266 & 1.197$\pm$0.252 & 20 & - & - & - & - & - & - \\
NGC~2257 & \textbf{1.400$\pm$0.025} & \textbf{0.958$\pm$0.033} & 121 & 1.413$\pm$0.100 & 0.949$\pm$0.138 & 252 & - & - & -  \\
%Hodge~3 & - & - & - & - & - & - \\
Hodge~4 & \textbf{1.632$\pm$0.080} & \textbf{0.379$\pm$0.090}& 68 & 1.566$\pm$0.084 & 0.342$\pm$0.093 & 13 & - & - & -  \\
Hodge~7 & 1.519$\pm$0.113 & 0.691$\pm$0.111 & 72 & -  & - & - & 1.579$\pm$0.01 & 0.709$\pm$0.01 & 949 \\
Hodge~11 & \textbf{1.466$\pm$0.034} & \textbf{0.989$\pm$0.049} & 110 & 1.468$\pm$0.165 & 1.007$\pm$0.220 & 95 & - & - & -  \\
%Hodge~14 & - & - & - & - & - & - \\
Reticulum & - & - & - & \textbf{1.964$\pm$0.093} & \textbf{-0.307$\pm$0.099} & 139  & 1.810$\pm$0.01 & -0.230$\pm$0.01 & 158  \\
\enddata
\tablecomments{All PM measurements are in ~mas~yr$^{-1}$. The bold entry indicates the bulk PM measurement used throughout the rest of the work. Those clusters without a bold entry, are not part of the final sample due to a lack of literature distance estimate and/or LoS velocity (see Table \ref{tab:prop}). }
\end{deluxetable*}

\subsubsection{Method 2} \label{subsec:M2}

In Method 2, which is described in more detail in Watkins et al. (in prep.), we assume that towards the center of the cluster there will be a mix of cluster and field stars, while away from the cluster there will be only field stars. 

Initially stars without PM measurements are removed. Stars with spurious measurements are also removed by selecting only stars with colors 10$>$$G_{BP}-G_{RP}$$>$-10 mag, proper motions 50$>$$\mu_{\alpha}$, $\mu_{\delta}$$>$-50 mas~yr$^{-1}$ and parallax 10$>$$\varpi$$>$-10 mas. 
Finally, any stars with a blending fraction greater than 0.2 are cut, as are stars which have poor photometry or astrometry by standard measures for \textit{Gaia} data \citep[see][for example]{Riello2021}. 
This creates a clean sample on which to run the membership selection method.

 At this stage, the sample contains a mix of cluster stars and field stars, and the challenge is to disentangle them. So to begin, we first define the properties of the field population. We do this by selecting stars in an annulus around the cluster. The annulus needs to be far enough away from the cluster centre that there are no more cluster stars present and the sample is purely field stars, but not so far away that the properties of the field stars are different from the population contaminating the cluster sample. 
We use visual selection to ensure the annulus is outside the overdensity of stars caused by the cluster.

Once a sample of field stars has been established we examine the stars in the vicinity of the cluster center. Here, we make use of three metrics to assess the likelihood of a star being a cluster member or field star. These are the position of the star in the color-magnitude diagram, the position of the star within the proper motion-parallax phase space and finally the spatial position of the star relative to the center of the cluster, with this final metric being weighted at 0.33 of the weight compared to the first two. 

These metrics are combined to determine the probability for cluster membership for each star within the region of the cluster. The stars with a membership probability $>$0.5 are then selected as cluster stars, while others are considered to be field stars. These stars are then iteratively fit with a series of 2D Gaussian profiles in proper motion space. These profiles determine a bulk proper motion and dispersion for the cluster, which is then used to cut outlying stars. This process is run again on the newly cut sample, this is repeated until no stars are cut. Those clusters which are found to have less than 10 high-quality cluster members are regarded as having unreliable fits, as are those clusters with a final dispersion $>$0.5 mas~yr$^{-1}$. 

Using this method we derive good bulk PMs for $18$ of the $42$ GCs from the examined catalogues. These proper motions are listed in Table~\ref{tab:PM}. The middle row of Figure~\ref{fig:NGC1783} illustrates the data and results for NGC~1783.

\subsubsection{GaiaHub} \label{subsec:gaiahub}

As a third method, we derive PMs by comparing the positions of stars in archival \textit{HST} GC imaging (as first epoch) with the positions of those same stars in the eDR3 (as second epoch). For this we use the GaiaHub\footnote{https://github.com/AndresdPM/GaiaHub} pipeline described in \citet{delPino2022}, which expands on and automates the approach previously used in e.g. \citet{Massari2017}. The advantage of this method is that it yields higher PM accuracy at faint magnitudes, which is especially important for distant GCs (or other dwarf galaxies) that lack bright young stars. The disadvantage is that it is limited to objects with suitable pre-existing archival imaging, and to the size of the \textit{HST} field of view. We run GaiaHub using the default settings except for enabling the \mbox{\tt --use\_members} flag, this flag means GaiaHub only uses the member stars to perform the alignment between the epochs, for more details see the appendix of \citet{delPino2022}. 

Of the 42 GCs examined, 12 were found to have \textit{HST} imaging compatible with this method. The inferred PMs are listed in Table~\ref{tab:PM}. The bottom row of Figure \ref{fig:NGC1783} illustrates the data and results for NGC~1783. For 6 of the clusters, the results support those already obtained with the previously described methods. But GaiaHub also yields bulk PMs for 6 clusters that were not measurable with the previously described methods.

\subsection{Proper Motion Comparison and Validation} \label{subsec:comp}

\begin{figure*}
\begin{center}
 \includegraphics[width=14cm]{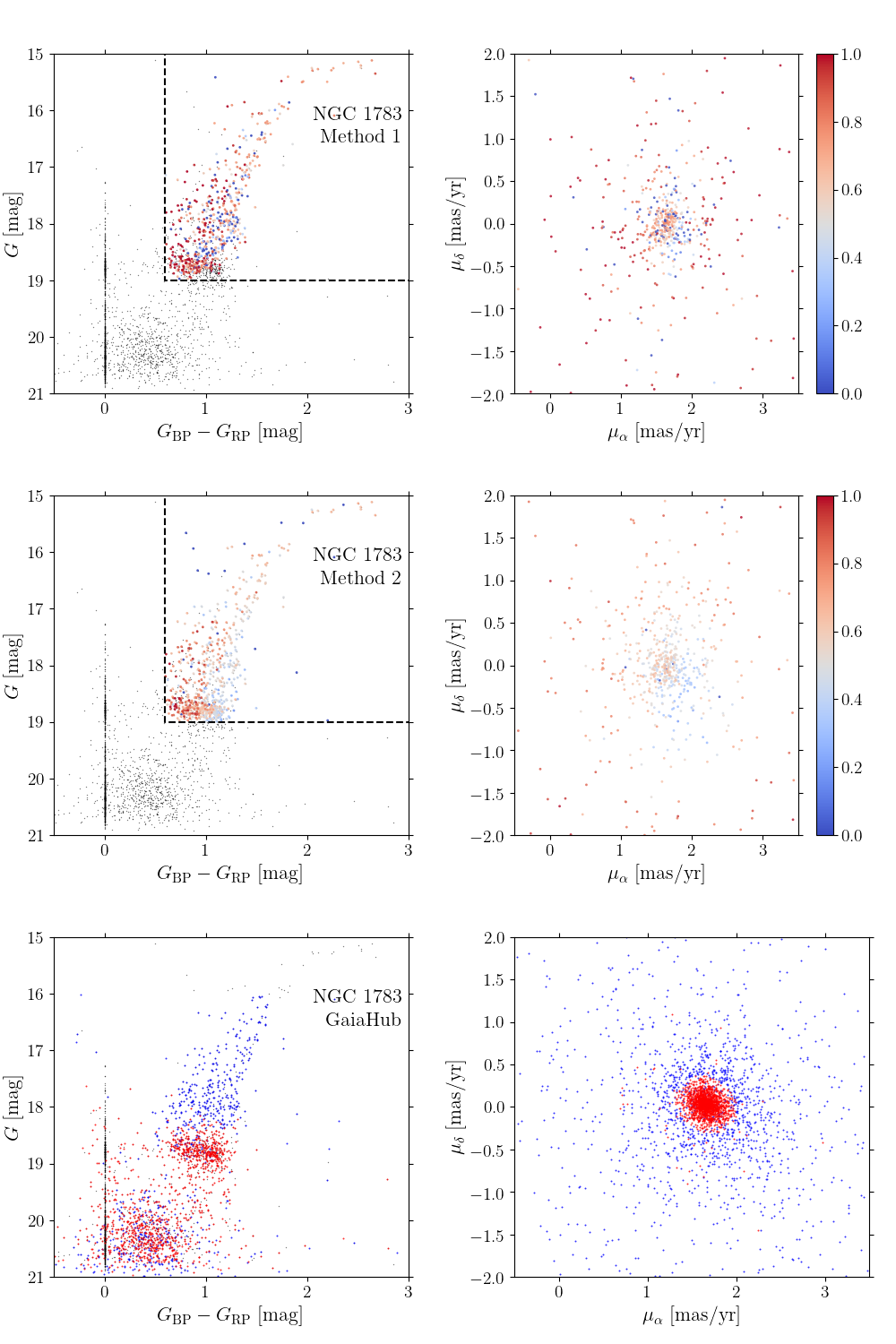}
 \caption{The CMD(left) and vector point diagram (right) for the GC NGC 1783 via all three methods used in this work (see \S \ref{sec:data}). Membership is shown by color with higher probability members in red and lower probability members in blue, membership probability is not shown for GaiaHub where all members stars are shown in red and background stars are shown in blue. Black points indicate Gaia sources removed by quality cuts. The magnitude and color based cuts for eDR3 based methods are shown as black dashed lines. Upper: Method one. Middle: Method 2. Lower: GaiaHub. }
 \label{fig:NGC1783}
 \end{center}
\end{figure*}

The different methods can be seen in Figure \ref{fig:NGC1783} applied to NGC 1783. This first demonstrates the similarity of the results reported by the two \textit{Gaia} eDR3 based methods in both the CMD and vector point diagram for the stellar proper motions. The red giant branch and red clump for NCG~1783 can be clearly seen in the CMD and the methods clearly distinguish these features from the equivalent features of the LMC's stellar disk, which can be seen in the redder lower-probability stars. 
The total number of high-probability member stars is higher in method 1 than method 2, with 426 for the former and 56 for the latter, despite this difference between the two methods we see strong agreement in the bulk PMs reported. 

We see a substantial difference in the reported error on the bulk PMs, with method 2 having reported almost 3 times higher errors than method 1. This is caused by the larger number of stars in method 1 as the error is inversely proportional to the square root of the number of stars. This difference in the number of high-probability GC member stars is likely due to the difference in how the methods deal with potential contamination from the LMC's stellar disk. Method 2 tends to be more cautious when considering stars that could be part of the LMC's stellar disk in either CMD or PM space, this leads to higher reported error as the number of high quality members is lower. 
This means that method 1 reports similar errors for all clusters, with smaller errors for those clusters with more high probability members. 
The number of reported members also effects the results from method 2, but method 2 also depends on cluster position with smaller reported errors for clusters where the background is less likely to contaminate the GC. In practise this means method 1 responds better to GCs that are projected over the LMC's stellar disk, whereas method 2 works best for clusters that are not.

From Figure \ref{fig:NGC1783} it can be seen that substantially more stars are available for analysis in GaiaHub (1782) than for either of the \textit{Gaia} eDR3 only methods. This is because the use of \textit{HST} means that the data for dimmer stars is more reliable than the same stars in \textit{Gaia} eDR3 alone. This large increase in the number of stars analyzed allows a substantial increase in the precision of the proper motion measurement, because the statistical errors scale inversely with the square root of the number of stars used. This increase in the number of stars and the ability to use fainter stars is also why GaiaHub is able to recover bulk PMs for 6 clusters where the \textit{Gaia} eDR3 methods can not distinguish the GCs from the LMC's stellar disk. 

Of the 42 GCs from the catalogs, 14 have PMs found by both \textit{Gaia} eDR3 methods, 8 are found only by method-1, 4 are found only by method-2 and 5 are found only in GaiaHub. This gives us a total sample of 31 GCs where we have derived a reliable PM measurement in this work. % from any method. 
The fact that these different methods were successful at deriving PMs for different subsets of clusters is intriguing and may have to do with how they characterize the background LMC population, but a detailed investigation of the cause of this difference is beyond the scope of this paper. 

\begin{figure*}
\begin{center}
 \includegraphics[width=16cm]{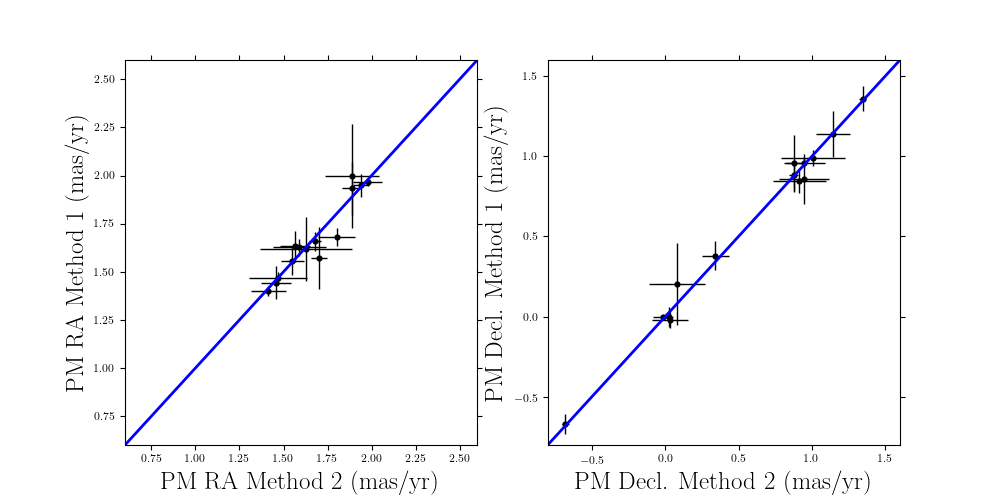}
 \includegraphics[width=16cm]{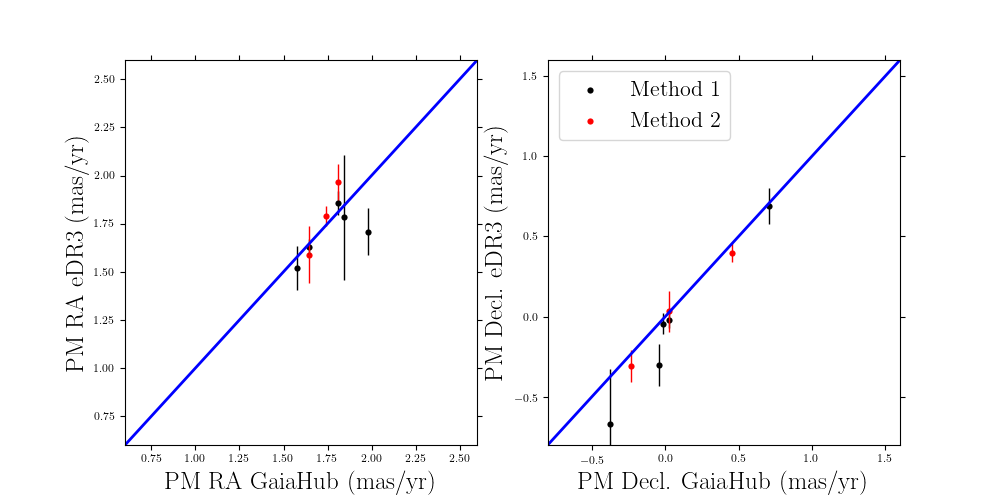} 
 \caption{Upper: Comparison between Method 1 and Method 2 for the derived proper motions for the 14 GCs where both methods produce good PM measurements. Left panel shows PM in R.A. and the right panel shows PM in Decl. Lower: Comparison between GaiaHub PMs and those from methods 1 and 2 for the 8 clusters where there are comparable PMs. The blue lines indicate a 1:1 agreement. All panels show strong agreement between the different methods. 
 \label{fig:comp}}
 \end{center}
\end{figure*}

We present a comparison of the proper motion values obtained with method~1 and method~2 in this work in Table \ref{tab:PM} and in the upper panels of Figure \ref{fig:comp}. We also compare the GaiaHub results to those obtained via method~1 or method~2 in Table \ref{tab:PM} and the lower panels of Figure \ref{fig:comp}. Despite the differences between the independent methods we have obtained broadly consistent results (within 1$\sigma$) for all GCs where more than one method could identify a proper motion. We find that the root mean square (RMS) of the differences between method 1 and method 2 is $\sim$0.055 mas~ yr$^{-1}$ which is substantially less than the typical observational uncertainties in the PM ($\sim$0.12 mas~yr$^{-1}$), this shows the methods are in agreement and any differences are caused by the observational uncertainties. Similarly the RMS between the eDR3 methods and GaiaHub is $\sim$0.12 mas~yr$^{-1}$, indicating agreement as this is consistent with the observational uncertainties. We have also compared the mean difference of the different methods and these were found to be consistent within 1 $\sigma$ for all tested coordinates and methods, indicating that any systematic bias is less than the observational uncertainties. 

For the rest of this work we will use a unified sample drawn from whichever of method~1 or method~2 produced the lowest uncertainty values on the bulk proper motion. This is generally,  but not universally, the method with the higher number of high-quality GC member stars. GaiaHub results are only used for the 6 GCs where they are the only option and are otherwise used to confirm the results from method~1 and method~2. The final adopted values are highlighted in Table \ref{tab:PM}.

As mentioned in \S \ref{sec:intro}, \citet{Piatti2019} presented the proper motions of a sample of 15 LMC GCs using \textit{Gaia} DR2 data. The data for these objects was cut in a similar way to this work, limiting the sample to stars with G$\leq$19 mag and proper motion errors $\leq$ 0.5 mas~yr$^{-1}$. We determined PMs for 9 of the 15 clusters, and generally found good agreement between the earlier DR2 measurements and (at least one of) the methods used on eDR3 data as part of this work.The RMS between the results in this work and those of \citet{Piatti2019} is $\sim$0.09 mas~yr$^{-1}$, below the stated observational uncertainties $\sim$0.10 mas~yr$^{-1}$. We also find the mean difference to be consistent within 1 $\sigma$. For the other 6 GCs which that do not have reliable PM measurements from the techniques used in this work, 
we use the PM results reported in \citet{Piatti2019}, and add these clusters (identified in the last column of Table~\ref{tab:vel}) to our sample in Table~\ref{tab:PM}. This creates a final sample of 36 clusters with PM measurements, of which 32 have full 6D information. 

\begin{deluxetable*}{c|ccc|cccc|c}
\tablenum{3}
\tablecaption{Globular clusters Velocities \label{tab:vel}}
\tablewidth{0pt}
\tablehead{ \colhead{Cluster} & \colhead{R} & \colhead{$\phi$} & \colhead{Z} & \colhead{V$_{R}$} & \colhead{V$_{\phi}$} & \colhead{V$_{\phi~resid.}$} & \colhead{V$_{z}$} & \colhead{PM}\\
\multicolumn1c{Name} & \multicolumn1c{(kpc)} & \multicolumn1c{(rad.)} & \multicolumn1c{(kpc)} & \multicolumn1c{(km~s$^{-1}$)} & \multicolumn1c{(km~s$^{-1}$)} & \multicolumn1c{(km~s$^{-1}$)} & \multicolumn1c{(km~s$^{-1}$)} & \multicolumn1c{Source} 
}
%\decimalcolnumbers
\startdata
 NGC1466 & 8.63 & 1.98 & -3.51 & 35$\pm$7 & 73$\pm$8 & 12$\pm$8 & 43$\pm$5 & M2 \\
 NGC1644 & 4.95 & 3.07 & 0.23 & 0$\pm$31 & 82$\pm$25 & 5$\pm$25 & 35$\pm$17 & M1 \\
 NGC1651 & 2.23 & 2.58 & 2.14 & -9$\pm$38 & 72$\pm$33 & 25$\pm$33 & 42$\pm$21 & M1 \\
 NGC1652 & 3.43 & 2.77 & 0.58 & -32$\pm$78 & 163$\pm$65 & 94$\pm$65 & 39$\pm$43 & M1 \\
 NGC1754 & 2.06 & 2.23 & 0.04 & -5$\pm$12 & 71$\pm$13 & 28$\pm$13 & 35$\pm$9 & P19 \\
 NGC1783 & 3.99 & 3.49 & 1.05 & 3$\pm$11 & 96$\pm$11 & 16$\pm$11 & 7$\pm$10 & M1 \\
 NGC1786 & 2.74 & 3.41 & 1.42 & -19$\pm$30 & 33$\pm$26 & -23$\pm$26 & -17$\pm$18 & M1 \\
 NGC1806 & 2.39 & 3.34 & 0.95 & -27$\pm$16 & 27$\pm$13 & -23$\pm$13 & 48$\pm$9 & M1 \\
 NGC1818 & 3.31 & 3.26 & -0.83 & 49$\pm$10 & 93$\pm$9 & 26$\pm$9 & -12$\pm$6 & M2 \\
 NGC1831 & 4.88 & 3.81 & 2.44 & -25$\pm$12 & 80$\pm$11 & 3$\pm$11 & 19$\pm$8 & M1 \\
 NGC1835 & 1.49 & 2.69 & 0.00 & -54$\pm$5 & 11$\pm$5 & -22$\pm$5 & 63$\pm$5 & P19 \\
 NGC1841 & 11.59 & 0.86 & -0.28 & 50$\pm$4 & 46$\pm$4 & -4$\pm$4 & 41$\pm$3 & M1 \\
 NGC1866 & 3.72 & 3.68 & 0.13 & -62$\pm$8 & 15$\pm$7 & -60$\pm$7 & -23$\pm$5 & M2 \\
 NGC1868 & 4.42 & 3.32 & -3.81 & 25$\pm$54 & 30$\pm$48 & -48$\pm$48 & -8$\pm$32 & M1 \\
 NGC1898 & 1.13 & 4.23 & 2.24 & 14$\pm$4 & -21$\pm$4 & -47$\pm$4 & 41$\pm$5 & P19 \\
 NGC1916 & 0.59 & 3.02 & -0.02 & 34$\pm$16 & -11$\pm$15 & -27$\pm$15 & -29$\pm$10 & P19 \\
 NGC1928 & 0.44 & 2.81 & -0.25 & -30$\pm$3 & -3$\pm$7 & -17$\pm$7 & 3$\pm$11 & GH \\
 NGC1939 & 0.11 & 3.21 & 0.44 & 39$\pm$2 & 8$\pm$5 & 1$\pm$5 & 4$\pm$6 & GH \\
 NGC1978 & 2.88 & 4.07 & 0.31 & 4$\pm$11 & 24$\pm$12 & -35$\pm$12 & -19$\pm$7 & M1 \\
 NGC2005 & 0.88 & 5.03 & 1.12 & 42$\pm$11 & 28$\pm$12 & 6$\pm$12 & -38$\pm$8 & P19 \\
 NGC2019 & 0.61 & 6.21 & -0.02 & 17$\pm$13 & -6$\pm$12 & -22$\pm$12 & -15$\pm$8 & P19 \\
 NGC2159 & 1.69 & 5.85 & -3.26 & 5$\pm$16 & 178$\pm$20 & 141$\pm$20 & 170$\pm$27 & M2 \\
 NGC2162 & 4.19 & 4.56 & -1.99 & 56$\pm$17 & 81$\pm$20 & 2$\pm$20 & -24$\pm$12 & M1 \\
 NGC2173 & 3.43 & 5.99 & 3.02 & 15$\pm$8 & 51$\pm$7 & -19$\pm$7 & 8$\pm$5 & M2 \\
 NGC2190 & 7.50 & 0.83 & -6.38 & -32$\pm$13 & 100$\pm$13 & 34$\pm$13 & -48$\pm$9 & M2 \\
 NGC2209 & 4.30 & 6.28 & 1.78 & 13$\pm$36 & 49$\pm$32 & -30$\pm$32 & -7$\pm$23 & M2 \\
 NGC2210 & 3.23 & 5.66 & -1.70 & -32$\pm$11 & 117$\pm$11 & 52$\pm$11 & -111$\pm$8 & M2 \\
 NGC2231 & 3.58 & 5.58 & -2.14 & 59$\pm$20 & 98$\pm$20 & 26$\pm$20 & -35$\pm$13 & M2 \\
 NGC2257 & 5.37 & 5.14 & -2.07 & 27$\pm$7 & 69$\pm$7 & -5$\pm$7 & -8$\pm$4 & M1 \\
 HODGE4 & 2.66 & 3.47 & -4.64 & 16$\pm$22 & 17$\pm$19 & -38$\pm$19 & -34$\pm$13 & M1 \\
 HODGE11 & 3.32 & 6.01 & -1.52 & -46$\pm$10 & 74$\pm$9 & 7$\pm$9 & 15$\pm$6 & M1 \\
Reticulum & 10.59 & 3.39 & -0.51 & -33$\pm$24 & 16$\pm$19 & -37$\pm$19 & -21$\pm$13 & M2 \\
 \enddata
\tablecomments{M1 is method 1 (see \S \ref{subsec:M1}), M2 is method 2 (see \S \ref{subsec:M2}), GH is GaiaHub (see \S \ref{subsec:gaiahub}), P19 is \citet{Piatti2019}}
\end{deluxetable*}

\begin{figure*}
\begin{center}
 \includegraphics[width=18cm]{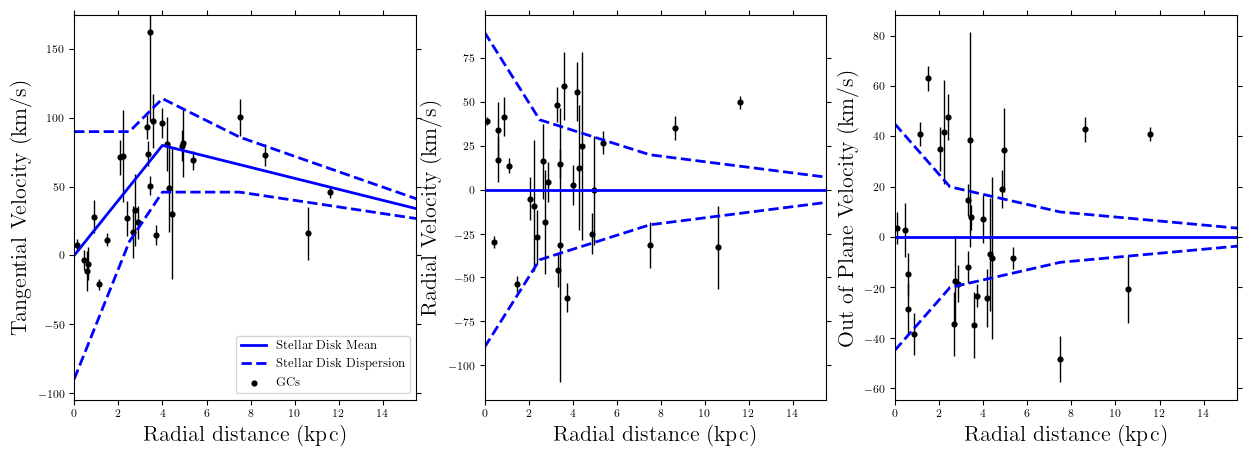}
 \caption{ GC velocities compared to radial distance from the center of the LMC. The black points are the GCs, the solid blue line indicates the mean velocity from the stellar disk and the dashed blue lines indicate the dispersion from the stellar disk \citep[stellar disk model is based on][]{Gaiacollab2021}. Left: Azimuthal velocity. Center: Radial Velocity. Right: Out-of-plane velocity. From these plots it can be seen that the GC population follows closely to the LMC's stellar population, with consistent velocities and expected numbers of clusters within the velocity dispersions. NGC~2159 and NGC~2210 are not shown in this plot, see \S \ref{subsec:outlier} for details. \label{fig:rot}}
 \end{center}
\end{figure*}

\subsection{Coordinate System and Velocity decomposition} \label{coordinates}

To examine the GCs in a physically meaningful way we need to change the reference frame from sky coordinates to a reference frame centered on the LMC. This requires use of spherical trigonometry, since the LMC covers a sufficiently large sky area that the common approximation that the target galaxy is at effectively infinite distance, and consequently that the ``sky is a flat plane", does not apply. To perform the coordinate transformations we use the prescription and equations of \citet{vanderMarel2001} and \citet{vanderMarel2002}. Starting from the sky position, distance, LoS velocity, and PM of a cluster, this yields the position and velocity in a Cartesian reference frame centered on the LMC and aligned such that the X-Y plane is coincident with the disk of the LMC, with the X-axis being along the line of nodes.

To determine coordinates relative to the LMC center and disk plane, it is necessary to specify the respective position, distance, velocity, PM, and geometry of those. We adopt an LMC center of ($\alpha_0,\delta_0)=(81.28\degree,-69.78\degree)$ \citep{vanderMarel2001}. We assume an LMC distance of $\mathrm{D_{LMC}}=50.1\pm2.5$~kpc \citep{vanderMarel2002}. We adopt an LMC LoS velocity of $262.2\pm3.4$~km~s$^{-1}$ \citep{McConnachie2012}, and proper motion \hbox{$(\mu_W,\mu_N)=(-1.910\pm0.020,0.229\pm0.047)$~mas~yr$^{-1}$} \citep{Kallivayalil2013}. We take the 
inclination angle $i$ of the disk to be $i=37.4\degree\pm6.2\degree$ \citet{vanderMarel2001} and the position angle of the line of nodes 
as $\Theta=-129\degree\pm8.3\degree$
\citet{vanderMarel2001}. For all of these quantities there exist many determinations in the literature obtained from a broad array of methods and tracer populations. However, the uncertainties in these quantities do not significantly affect the subsequent discussion. The scatter in the GC kinematics are dominated by the internal velocity dispersion of the system and the PM uncertainties, and not uncertainties in the geometry of the adopted Cartesian system.

Once we have obtained the GC coordinates and velocities in the aforementioned reference frame, we transform the 3D velocities in this frame into spherical and cylindrical polar coordinates. These velocity components are particularly useful for understanding the dynamics of the cluster system. Specifically, for the remainder of this paper we
we will focus on positions and velocities in cylindrical coordinates. Using the data in Tables~\ref{tab:prop} and \ref{tab:PM}, combined with the PMs for the 6 additional clusters from \citet{Piatti2019}, this yields results for 32 GCs. These are listed in Table~\ref{tab:vel}. We note that there are four GCs with PMs that do not have a literature determination of either the LoS velocity or distance. An expansion of the sample to include these targets would be a useful addition to this work for relatively little time in additional observational follow-up.

%########################################################

\begin{deluxetable*}{c|cc|cc|cc|c}
\tablenum{4}
\tablecaption{Velocity mean and dispersion \label{tab:disp}}
\tablewidth{0pt}
\tablehead{
\colhead{} & \colhead{$\bar{V_{R}}$} & \colhead{$\sigma_{R}$} & \colhead{$\bar{V_{\phi}}$} & \colhead{$\sigma_{\phi}$} & \colhead{$\bar{V_{Z}}$} & \colhead{$\sigma_{Z}$}  \\
\multicolumn1c{} & \multicolumn1c{(km~s$^{-1}$)} & \multicolumn1c{(km~s$^{-1}$)} & \multicolumn1c{(km~s$^{-1}$)} & \multicolumn1c{(km~s$^{-1}$)} & \multicolumn1c{(km~s$^{-1}$)} & \multicolumn1c{(km~s$^{-1}$)} 
}
%\decimalcolnumbers
\startdata
Literature distances & 5.02$\pm$6.96 & 33.78$\pm$5.40 & 8.22$\pm$4.91 & 22.50$\pm$4.90 & 3.22$\pm$5.83 & 29.58$\pm$4.49  \\
Bootstrap distances & 1.71$\pm$6.42 & 31.05$\pm$5.08 & 3.13$\pm$6.19 & 29.93$\pm$4.91 & -1.14$\pm$6.09 & 31.09$\pm$4.67  \\
Disk Distances & -3.85$\pm$5.77 & 27.22$\pm$4.59 & 1.98$\pm$6.32 & 31.36$\pm$5.09 & 3.93$\pm$4.81 & 23.54$\pm$3.73  \\
%Monte Carlo & 0.30$\pm$5.80 & 24.65$\pm$5.32 & 53.98$\pm$6.79 & 31.14$\pm$5.82 & 0.06$\pm$5.49 & 26.79$\pm$4.46 & 0.20$\pm$0.43\\
\enddata
%\tablecomments{Globular...}
\end{deluxetable*}

\begin{deluxetable*}{c|cc|cc|cc}
\tablenum{5}
\tablecaption{Velocity mean and dispersion for sub-populations \label{tab:disp_sub}}
\tablewidth{0pt}
\tablehead{
\colhead{Sub-population} & \colhead{$\bar{V_{R}}$} & \colhead{$\sigma_{R}$} & \colhead{$\bar{V_{\phi}}$} & \colhead{$\sigma_{\phi}$} & \colhead{$\bar{V_{Z}}$} & \colhead{$\sigma_{Z}$} \\
\multicolumn1c{} & \multicolumn1c{(km~s$^{-1}$)} & \multicolumn1c{(km~s$^{-1}$)} & \multicolumn1c{(km~s$^{-1}$)} & \multicolumn1c{(km~s$^{-1}$)} & \multicolumn1c{(km~s$^{-1}$)} & \multicolumn1c{(km~s$^{-1}$)}
}
%\decimalcolnumbers
\startdata
 Log(Age)$\leq$9.5 (yrs) & 5.77$\pm$10.90 & 37.12$\pm$8.54 & 5.90$\pm$4.39 & 12.21$\pm$4.02 & 7.00$\pm$9.08 & 31.39$\pm$7.03 \\
 Log(Age)$>$9.5 (yrs)& 4.15$\pm$9.35 & 32.26$\pm$7.71 & 8.90$\pm$8.62 & 29.58$\pm$7.09 & -0.14$\pm$7.83 & 29.14$\pm$6.29 \\
 $M_{V}\leq-7.5$ & 0.47$\pm$9.62 & 36.05$\pm$7.54 & 16.10$\pm$6.87 & 24.32$\pm$5.54 & 8.35$\pm$8.04 & 30.74$\pm$6.22 \\
 $M_{V}>-7.5$ & 11.74$\pm$10.18 & 31.59$\pm$8.33 & -1.74$\pm$6.31 & 16.36$\pm$5.41 & -4.35$\pm$8.46 & 27.38$\pm$6.88 \\
 Log(Mass)$\leq$4.8 (M$_{\odot}$)  & 1.69$\pm$22.31 & 50.74$\pm$18.87 & -1.14$\pm$18.25 & 41.12$\pm$15.67 & -14.5$\pm$11.31 & 25.12$\pm$9.78 \\
 Log(Mass)$>$4.8 (M$_{\odot}$) & 3.68$\pm$10.10 & 30.45$\pm$8.56 & 6.34$\pm$6.01 & 15.19$\pm$5.03 & 15.59$\pm$8.33 & 25.57$\pm$6.87 \\
\enddata
\tablecomments{Metallicity has been found to closely correlate with age in LMC clusters \citep[e.g.][]{Harris_2009}, therefore age is used as a proxy from metallicity as many of the sample GCs do not have published metallicity estimates. 
Mass estimates from \cite{Baumgardt2013} are used, these cover 19 of the 30 GCs in the final sample, therefore this does not use the full sample, this reduced sample is also biased in favor of young clusters as 10 of the missing 12 clusters are old (Log(Age)$>$9.5 yrs). }
\end{deluxetable*}

\begin{figure*}
\begin{center}
 \includegraphics[width=18cm]{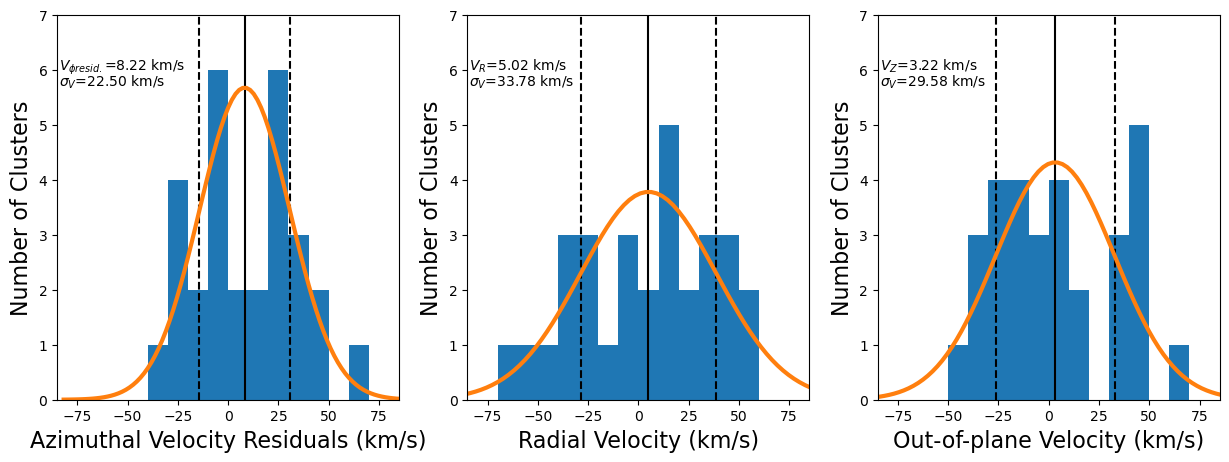}
 \caption{ Histograms plotting the cylindrical coordinate velocities of the GCs. The solid black line marks the mean value and the dashed lines indicate the dispersion assuming a Gaussian profile. All three distributions shown are statistically indistinguishable from Gaussian distributions at a $p<0.10$, these Gaussian distributions are plotted in solid orange lines. The anisotropy is indirectly apparent as the velocity dispersion of the azimuthal coordinate is substantially lower than that seen in the other coordinates. 
 \label{fig:disp}}
 \end{center}
\end{figure*}

\section{Kinematics of the LMC GC system} \label{sec:result}

Figure \ref{fig:rot} shows the cylindrical azimuthal, radial, and vertical velocity components of the GCs as a function of cylindrical radius. NGC 2159 and NGC 2210 fall outside of the vertical boundaries of the plot, and were excluded from the subsequent analysis for the reasons discussed in Section~\ref{subsec:outlier} below. 

We compare the observed azimuthal velocities of the GCs to the rotation curve derived for the stellar populations of the LMC (shown as blue lines). 
Current best-fit models for the stellar rotation of the LMC from \textit{Gaia} eDR3 show a rotation curve that rises from 0 km~s$^{-1}$ at the center of the LMC to 80 km~s$^{-1}$ at 4.5 kpc, before falling to 64 km~s$^{-1}$ around the edge of the LMC disk at 8 kpc \citep[][]{Gaiacollab2021}. As an approximation, we continue this trend in the figure out to the radius of the most distant GC ($\sim$12 kpc). 

The LMC GCs have fairly well-behaved kinematics and appear to be orbiting the LMC in a manner consistent with the stellar disk\footnote{Given that the angular size of the LMC is larger than the scale of Gaia's systematic errors \citep[][]{Lindegren2021} we need to consider if these will have an effect on our observations. However we find that the scale of these systematics is far smaller than the dispersion between our clusters and the statistical uncertainty in the PM for individual clusters. Therefore we consider these effects to be negligible for the purpose of this paper.}. We also see support for several features found in the LMC's stellar population, such as counter-rotation in the LMC's core near the bar \citep[][]{Gaiacollab2021}, with 4 of the innermost 6 GCs exhibiting counter-rotation. This includes NGC~2005 which was recently speculated to be part of a small dwarf accreted by the LMC \citep[][]{Mucciarelli2021}, indicating there may be other clusters that were part of this accreted dwarf. Though this observed counter rotation in the GCs is possibly due to distance uncertainties. These uncertainties could shift clusters close to the center to the `wrong' side of the LMC causing them to appear to counter rotate. 

To analyze the velocity dispersions of the GC system, we take the mean azimuthal velocity of the underlying population to be that of the stellar rotation curve, and the mean radial and vertical velocities to be zero. The residuals with respect to these adopted means are shown in Figure~\ref{fig:disp}. 
We find that a Gaussian distribution can fit the velocity distribution in all coordinates, these distributions can been seen in Figure \ref{fig:disp}. The Anderson-Darling test \citep[][]{Anderson1952} found that a Gaussian distribution could not be ruled out at p$<$0.10 in all three coordinates. However, the statistical power of this, and other statistical tests, is limited by the small number of total clusters and thus many types of distribution cannot be ruled out. We discuss the possibility of the existence of multiple populations in the GC distribution in Section~\ref{sec:pops}. 

The average and dispersion of the residuals are listed in the top row of Table~\ref{tab:disp}. These were calculated using a bootstrap repeated 10,000 times with the proper motions and LoS velocity varied within a Gaussian distribution based on the reported errors for each GC. The averages are generally consistent with zero, justifying our choice of the mean for the underlying population (there is a mild indication for the GCs to rotate faster than the stars, but not at compelling significance).

The average one-dimensional velocity dispersion is 29 km~s$^{-1}$, consistent with that previously derived from LoS velocities alone \citep[][]{Schommer1992, Grocholski2006}. This dispersion is similar to that of old stellar populations in the LMC disk \citep[][]{vanderMarel2009}. As shown in Table \ref{tab:disp}, the velocity dispersions are not the same in all coordinates; they show a significant anisotropy. 
There is a higher dispersion and, therefore, higher random motions in the radial and out-of-plane directions compared to the azimuthal motion, once the overall rotation of the LMC is removed. We also find a slight difference between the radial and out-of-plane dispersions, though this difference is less pronounced and not statistically significant.

This anisotropy matches what would be expected from a system with organized disk-like rotation. For example, the MW disk GCs show a lower azimuthal velocity dispersion compared to the other coordinates \citep[][]{ Posti2019}. Also it is unlike the GCs of the MW halo, which show a rotation much slower than the disk, and a different anisotropy, with comparable dispersions in the radial and azimuthal coordinates ($\sim$120 km s$^{-1}$) and a slightly lower dispersion in the out-of-plane velocity ($\sim$100 km s$^{-1}$) \citep[calculated using the GC sample from][]{Watkins2019}. We can test this using the beta parameter. 

\begin{equation}
    \beta = 1 - \frac{(\sigma^{2}_{R}+\sigma^{2}_{z})}{(2\sigma^{2}_{\phi})}
\end{equation}

If the $\beta$ parameter should be zero for an isotropic dispersion, a positive $\beta$ shows relatively high azimuthal dispersion, whereas a negative $\beta$ shows a low azimuthal dispersion indicating ordered rotation. The $\beta$ parameter for the GCs is $-0.99^{+0.72}_{-1.42}$. On initial inspection this large error range seems to show that the negative value of $\beta$ is not significant, however the high asymmetry in the uncertainty of $\beta$ means that a positive value for $\beta$ is excluded at 94\% confidence, meaning we see significant evidence of ordered rotation.

To account for errors on the reported distances, we have rerun our previous analysis over 500 times with the distances randomized in Gaussian distributions based on the reported errors using a bootstrap method. This produced consistent velocity means and dispersions ($\sim$30 km~s$^{-1}$) across all 3 coordinates (see second row of Table \ref{tab:disp}), implying an isotropic velocity distribution of the residuals. This suggests that the observed anisotropy could be an artifact of observational uncertainties. On the other hand, isotropic velocity distributions are not observed in other rotational dominated cluster populations, like the MW disk clusters \citep[][]{ Posti2019}. 

We have also run a Monte Carlo test with 500 iterations. This uses a synthetic population of GCs with the velocities of the individual clusters assigned isotropically based on a mean 1D velocity of 30 km~s$^{-1}$ across all 3 coordinates and then a typical observational error added. In this test we find that the isotropic velocities show on average $\beta=-0.02\pm0.02$, where the uncertainty is the error in the mean. This is consistent with zero as expected, and is substantially higher than the real data which has a value of $-0.99^{+0.72}_{-1.42}$ indicating that the real data has lower azimuthal dispersion than can be explained by an isotropic distribution and that there is an ordered rotation.

\section{Discussion} \label{sec:discussion}

\subsection{Single versus Multiple GC Populations} 
\label{sec:pops}

It is clear from the preceding discussion, and especially Figure~\ref{fig:rot}, that GCs in the LMC form a population with predominantly disk-like kinematics. But it is important to assess whether there is any evidence for a secondary population. In the MW, the GCs fall into two distinct populations, a disk population and a halo population. 
These are believed to relate to two 
distinct formation mechanisms, with the halo clusters being accreted from historic mergers \citep[][]{Brodie2006} as opposed to being formed in the MW like the disk clusters. So we are specifically interested to assess whether there may be evidence for a halo population in the LMC as well. 

A scenario where the LMC's GCs are a single population orbiting approximately with the LMC's stellar disk has been previously favored by LoS velocity measurements \citep[][]{Schommer1992,Grocholski2006} and simulations \citep[][]{Goodwin1997}. Our findings are fully consistent with this. Our new measurements are well fit by a single population with kinematics similar to that of the stellar disk, both in mean velocity and in velocity dispersion, to within the uncertainties.  

We find that the measurements are well fit by a single Gaussian in all three coordinates. Using a Gaussian as the null hypothesis, this hypothesis could not be rejected at the $p<0.10$ using the Anderson-Darling test \citep[][]{Anderson1952} across all three cylindrical coordinates. The tested Gaussian distributions match those that fit the stellar disk. 

In spite of the inconclusive result of this test, we suggest that the LMC's GC population should share a single common origin, likely forming in the LMC's disk from the collapse of massive HI clouds \citep[][]{Kruijssen2014, Kruijssen2015}. Exactly why the LMC 
GC population would lack the kinds of halo clusters that are present in the MW is beyond the scope of this paper. However, it is likely related to the difference in assembly history or mass of these galaxies, from 1.38$\times$$10^{11}$ for the LMC to 1.2$\times$$10^{13}$ M$_{\odot}$ for the MW \citep[][]{Erkal2019, Watkins2019, Posti2019}. 

To assess the possible presence of a secondary population, we looked for substructure in the histograms of residual velocity in Figure~\ref{fig:disp}. The azimuthal velocity histogram can in principle be well-fit by the sum of two Gaussians, one that is rotating slower than the stellar disk and the other that is rotating faster. The former could be naturally interpreted as a halo population. However, the latter has no meaningful physical interpretation, since it is not possible for a population to rotate faster than the disk. Hence, shot noise is a more likely explanation for the potentially bi-modal structure of this histogram.

\citet{Piatti2019} concluded from their study of \textit{Gaia} DR2 PMs of 15 LMC GCs that 10 are disk clusters and 5 are halo clusters. This distinction was made on the basis of an out-of-plane velocity cut. However, a hard cut is not physically motivated. Even in a single disk population there will be clusters with vertical velocities out to 2-3$\sigma$, and even in a halo population there will be clusters of low vertical velocity. Observational uncertainties further exacerbate this. Instead, in our sample, the vertical velocity distribution is well-fit by a single Gaussian. 

There is a wide difference in physical properties such as metallicity and age between the LMC and MW GCs \citep[][]{Olszewski1991, Harris1996, Mackey2003}. In general we find that the LMC's globular cluster population is well-mixed in terms of physical properties, unlike the MW where the halo clusters are often younger and more metal-poor than the disk clusters which are old and comparatively metal-rich \citep[][]{Brodie2006}. 
So as a further investigation, we grouped the LMC GCs by physical properties such as mass and metallicity, and calculated their kinematics. All tested sub-populations retain similar kinematics to the population as a whole (see Table \ref{tab:disp_sub}). The only exception to this is age, where young clusters (log(age)$<$9.5) show a smaller azimuthal velocity dispersion than old clusters (log(age)$>$9.5). Both old and young clusters show the same anisotropy pattern and consistent mean velocities. Therefore we conclude that the old and young cluster populations are both disk clusters, but the different formation times give the young clusters less time to be disturbed by galaxy interactions and mergers and thus form a tighter correlation with the stellar disk than the old clusters. The only other potential anomaly is the mean azimuthal velocity being smaller for faint clusters compared to bright ones, however on review this seems to be an artifact of small number statistics when examining sub-populations and is consistent within 2$\sigma$. Other kinematic differences between sub-populations follow those that would be expected from a single disk-like population with a tilted disk. 
So this too provides no significant evidence for the existence of a distinct halo GC population in the LMC.

\subsection{GC Spatial Distribution}\label{subsec:spatial}

\begin{figure*}
\begin{center}
 \includegraphics[width=18cm]{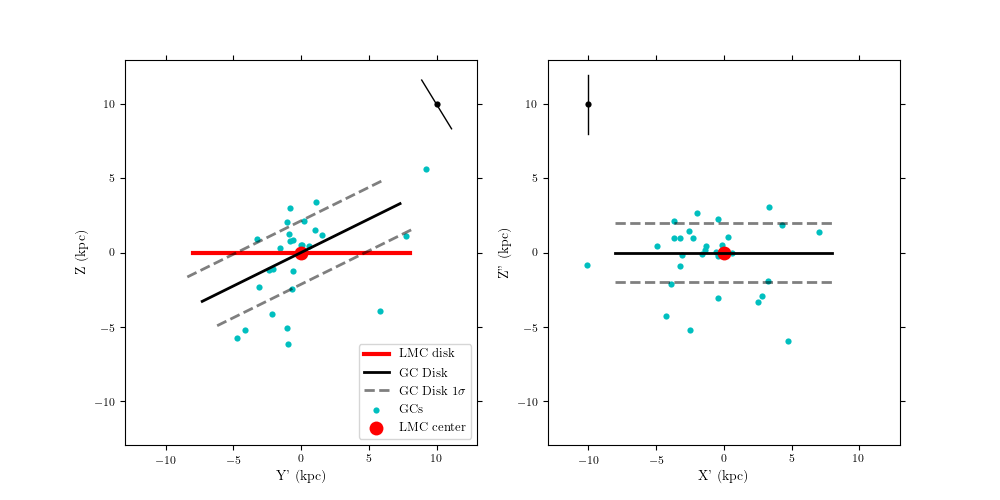}
 \caption{ Scatter plots illustrating the possible GC plane in the LMC, the cyan dots show the positions of the individual GCs, the solid black line illustrates the best fit of the GC plane, with the gray dotted lines show the 1$\sigma$ scatter based on the distance errors and does not take into account any intrinsic scatter in the GC distribution. The red line shows the LMC's stellar disk with the red dot showing the center of the LMC. The black dots show the mean distance error on the GCs. Left: Looking along the GC plane's line-of-nodes offset by 3.3 degrees from the LMC's line-of-nodes. Right: Looking along the GC plane having rotated about the GC's line-of-nodes (X' axis) by 24.1 degrees. 
 \label{fig:pos_2D}}
 \end{center}
\end{figure*}

Figure \ref{fig:pos_2D} illustrates the positions of our sample clusters as viewed edge-on from within the disk plane. As expected given their kinematics, the GCs occupy a near-planar distribution. This too argues against a significant halo population. But surprisingly, this disk-like structure is tilted with respect to the LMC stellar disk itself, and aligned partway between the LMC disk and the sky plane. The crossover between the LMC disk and the GC disk plane appears close to the LMC's line of nodes.

We performed a best fit of the GC distribution to a 2D plane based on a least squares approach. This produced a plane with a line-of-nodes that is offset by 3.3 degrees from the LMC's own, and a plane that is inclined by 24.1 degrees relative to the LMC's stellar disk. This plane is shown in Fig. \ref{fig:pos_2D}.  
As an additional test of this plane we have also examined the mean angle between the fitted GC spatial plane and the velocity vectors of the GC's, and find a mean angle of -0.03$\pm$0.09 radians. As this is consistent with zero within 1$\sigma$, and the $\beta$ parameter shows an anisotropy consistent with ordered rotation, we can conclude that both the distribution and kinematics of the GCs are consistent with this symmetry plane.

Investigations of the inclination of the LMC disk from a variety of tracers and methods have yielded results that span a significant range (10-20~degrees), well in excess of the statistical uncertainties\citep[][]{vanderMarel2009, Gaiacollab2021}. Also, variations in inclination within the LMC have been reported \citep[][]{vanderMarel2001,Choi2018}. Therefore, tilts between tracer populations as well as warps may well exist. However, tilts do not in general have a well-understood physical origin. Also in the present case, we did not identify an alignment with the direction towards the MW, the SMC, or the Magellanic stream that might explain its existence due to tidal interactions between the two galaxies.

Alternatively, it is possible that the apparent tilt may be the result of errors in the literature distances of the sample GCs. If there is a bias that places the presumed LMC clusters closer to the canonical distance of the LMC, then this could create the misaligned disk structure. This would explain why the crossover is closely aligned with the LMC's line of nodes (with an offset of only 3.3 degrees). As the line of nodes is a coordinate convention and has no physical origin, this alignment would be unexpected if there were a true intrinsic tilt.

In view of this, we also explored an alternative approach in which we fix the distances of all the GCs such that they are at the distance of the LMC's disk at the GC's sky coordinates using the model from \citep[][]{vanderMarel2002}. This is similar to the approach adopted 
by \citet{Piatti2019} for most of their sample clusters. The kinematics thus recalculated are reported in the third row of Table~\ref{tab:disp}.
While individual clusters have different properties, the properties of the GC population as a whole are very similar. However, the velocity anisotropy is reversed, with the azimuthal velocity dispersion being higher than the radial and out-of-plane velocity dispersions, with the latter two being comparable. This is the opposite of the anistropy observed for the MW disk cluster system, so this alternative approach also poses challenges in terms of physical interpretation. Either way, none of the preceding arguments in \S\ref{sec:pops} in favor of a single population were found to be meaningfully altered with these alternative distances.

\subsection{Outliers} 
\label{subsec:outlier}

NGC~2159 and NGC~2210 do not appear to follow the general trends of overall GC population, and they were excluded from the preceeding analyses. But while the focus of our discussion has been the GC system as a whole, it is also important to critically examine any potential outliers. 

NGC~2159 shows a fast orbital rotation and large out-of-plane velocity relative to the LMC or the other GCs in our sample (see Table~\ref{tab:vel}). Its relative velocity ($\sim$250 km~s$^{-1}$), is larger than the escape velocity of the LMC \citep[$\sim$170 km~s$^{-1}$ at the radius of NGC~2159,][]{Boubert_2017}. Therefore, we conclude that despite being physically close to the LMC, NGC~2159 cannot be a long-term member of the LMC system. Instead, it could be a Milky Way cluster that is currently near the LMC, or a cluster native to the LMC or accreted by the LMC from the SMC or an UFD, that is now being ejected due to an interaction. 
This interpretation may be supported by the relatively low radial velocity which would imply it is near the periapsis of a hyperbolic orbit. 
NGC~2159 is a young metal-rich cluster \citep[][]{Mackey2003}, making it unlikely to be a MW halo cluster, which are usually old and metal-poor \citep[][]{Harris1996}. This increases the likelihood that it is a cluster that is native to the Magellanic system, but is in the process of being ejected following an encounter between the LMC and another galaxy, probably the SMC. 

NGC~2210 also demonstrates unusual behavior with an extreme out-of-plane velocity and high general velocity ($\sim$160 km~s$^{-1}$). This is barely consistent with the escape velocity of the LMC at its radius ($\sim$170 km~s$^{-1}$). NGC~2210 is also an old metal-poor cluster \citep[][]{Olszewski1991} consistent with those of the MW halo \citep[][]{Harris1996}. NGC~2210 could therefore be a MW cluster that is either not physically associated with the LMC, or is in the process of being accreted by it.

These outliers also rely on the LoS velocity from literature. For these clusters this data is relatively sparse relying on only a few stars. Therefore, it is possible that their outlier status is caused by their LoS velocities being incorrect.

\section{Conclusions} \label{sec:conclusion}

We derived the bulk PMs of 31 LMC globular clusters. This doubles the sample size and increases the PM accuracy compared to previous studies. We used two independent methods using the \textit{Gaia} eDR3 data, as well as a software pipeline, GaiaHub, that derives PMs from comparison of positions in \textit{HST} archival data and the eDR3.

We combine the new PMs with literature values for distances, LoS velocities and existing PMs, to construct full 6D position-velocity information for 32 LMC clusters. Two of the GCs (NGC~2159 and NGC~2210) have highly discrepant velocities, suggesting that they are not long-term members of the LMC system. 

The kinematics of the rest of the sample are consistent with a single population rotating with the LMC's stellar disk. The one-dimensional velocity dispersions are of order 30 km~s$^{-1}$, similar to that of old stellar populations in the LMC disk. The azimuthal velocity has the lowest dispersion with the radial and vertical dispersions being higher. 
This is similar to the anisotropy of the MW disk clusters \citep[][]{Posti2019}. 
However, given the distance uncertainties, an isotropic distribution cannot be ruled out. In fact, the anisotopy could even be preferentially azimuthal if the literature distances are systematically in error. There is some evidence for this from the fact that while the GCs are configured in a disk-like structure, this structure appears tilted with respect to the stellar disk. 

We do not find evidence for multiple sub-populations within the GC system, or for a subset of clusters with halo-like kinematics. However, given the small number statistics, such a possible population cannot be ruled out completely. If there is indeed a single kinematic population of GCs in the LMC, then this strongly indicates that they have a single formation mechanism that has worked over the range of ages we see in the LMC's GC population from $\sim$10 Gyrs to $\sim$100 Myrs \citep[][]{Baumgardt2013}. This is in contrast to the MW where GCs are uniformly old, and fall into distinct disk and halo populations believed to correspond to different formation mechanisms.

\begin{acknowledgments}

Support for this work was provided by a grant for \textit{HST} archival program AR-15633 provided by the Space Telescope Science Institute, which is operated by AURA, Inc., under NASA contract NAS 5-26555.

A. del Pino acknowledge the financial support from the Spanish Ministry of Science and Innovation and the European Union - NextGenerationEU through the Recovery and Resilience Facility project ICTS-MRR-2021-03-CEFCA.

This work has made use of data from the European Space Agency (ESA) mission {\it Gaia} (\url{https://www.cosmos.esa.int/gaia}), processed by the {\it Gaia} Data Processing and Analysis Consortium (DPAC, \url{https://www.cosmos.esa.int/web/gaia/dpac/consortium}). Funding for the DPAC has been provided by national institutions, in particular the institutions participating in the {\it Gaia} Multilateral Agreement.

This research made use of Astropy,\footnote{https://www.astropy.org} a community-developed core Python package for Astronomy \citep{astropy2013, astropy2018}.

This research has made use of NASA's Astrophysics Data System Bibliographic Services.

This project is part of the HSTPROMO (High-resolution Space Telescope PROper MOtion) Collaboration\footnote{\url{http://www.stsci.edu/~marel/hstpromo.html}}, a set of projects aimed at improving our dynamical understanding of stars, clusters and galaxies in the nearby Universe through measurement and interpretation of proper motions from \textit{HST}, \textit{Gaia}, and other space observatories. We thank the collaboration members for the sharing of their ideas and software.

\end{acknowledgments}

\vspace{5mm}
\facilities{HST(ACS), Gaia }

\software{ AstroPy \citep[][]{astropy2013,astropy2018},  NumPy \citep[][]{harris2020}, SciPy\citep[][]{scipy2020} 
}

\bibliography{references}{}
\bibliographystyle{aasjournal}

\end{document}